\RequirePackage{fix-cm}

\documentclass[smallextended]{svjour3}       
\smartqed

\usepackage{graphicx}
\usepackage{times}
\usepackage[english]{babel}
\usepackage{amsmath}
\usepackage{siunitx}
\usepackage{amssymb}
\usepackage{subfig}
\usepackage{bm}
\usepackage{multirow}

\journalname{Meccanica}

\begin{document}

\title{Effects of atrial fibrillation on the arterial fluid dynamics: a modelling perspective}


\author{Stefania Scarsoglio \and Caterina Gallo \and Luca Ridolfi}

\institute{S. Scarsoglio \and
           C. Gallo \at Department of Mechanical and Aerospace Engineering, Politecnico di Torino, Torino, Italy\\  email{: stefania.scarsoglio@polito.it}                    \at
               \and L. Ridolfi \at Department of Environmental, Land and Infrastructure Engineering, Politecnico di Torino, Torino, Italy }

\date{Received: date / Accepted: date}

\maketitle

\begin{abstract}

Atrial fibrillation (AF) is the most common form of arrhythmia with accelerated and irregular heart rate (HR), leading to both heart failure and stroke and being responsible for an increase in cardiovascular morbidity and mortality. In spite of its importance, the direct effects of AF on the arterial hemodynamic patterns are not completely known to date. Based on a multiscale modelling approach, the proposed work investigates the effects of AF on the local arterial fluid dynamics. AF and normal sinus rhythm (NSR) conditions are simulated extracting 2000 $\mathrm{RR}$ heartbeats and comparing the most relevant cardiac and vascular parameters at the same HR (75 bpm). Present outcomes evidence that the arterial system is not able to completely absorb the AF-induced variability, which can be even amplified towards the peripheral circulation. AF is also able to locally alter the wave dynamics, by modifying the interplay between forward and backward signals. The sole heart rhythm variation (i.e., from NSR to AF) promotes an alteration of the regular dynamics at the arterial level which, in terms of pressure and peripheral perfusion, suggests a modification of the physiological phenomena ruled by periodicity (e.g., regular organ perfusion)and a possible vascular dysfunction due to the prolonged exposure to irregular and extreme values. The present study represents a first modeling approach to characterize the variability of arterial hemodynamics in presence of AF, which surely deserves further clinical investigation.

\end{abstract}

\section{Introduction}
\label{sec:1}
Atrial fibrillation (AF), with an estimated number of 33.5 million individuals affected worldwide in 2010, along with rising incidence, prevalence and mortality \cite{Chugh}, is the most wide\-spread cardiac arrhythmia. AF presents a disorganized electrical activity leading to irregular, uncorrelated and faster heart rate \cite{January}, and can yield significant symptoms, such as palpitations, chest pain, shortness of breath, weakness and reduced exercise ability \cite{January}, with an increased risk of stroke and heart failure \cite{Wolf,Roger} and a decreased quality of life.

Several clinical studies agree in finding a reduction of the cardiac efficiency and performance, in terms of reduced stroke volume and cardiac output during AF \cite{clark,daoud,alboni,upshaw,giglioli}. Although the impact of AF on the cardiovascular system is a subject of great medical interest, for the most part of the hemodynamic variables, such as cardiac and systemic arterial pressures, literature results are quite surprisingly contrasting \cite{clark,alboni,giglioli,kaliujnaya}. In vivo measurements face, indeed, a variety of different issues, which can hinder the observed hemodynamic scenario. For example, AF often arises with other concomitant pathologies (e.g., hypertension, atrial dilatation, valvular disorders, etc..), thus the net role of the arrythmia is not easily detectable. Moreover, oscillometric instruments do not work accurately in AF mainly because of the heart rate fluctuations \cite{Stergiou,Pagonas}, while current clinical techniques, such as Doppler echocardiography, do not have the resolving power to detail the instantaneous pressure and flow rate signals along the arterial districts \cite{Tae-Seok}. In addition, the variable cycle length induced by the cardiac disorganization requires suitable approaches, such as the wave intensity analysis \cite{Parker}, to estimate reflection and waveform indexes.

In order to overcome the current clinical lackings and become an effective supportive tool, computational hemodynamics has recently started to be adopted from a translational medicine perspective \cite{severi1,severi2}. In par\-ti\-cu\-lar, lumped-parameter approaches have been recently exploited to simulate AF conditions and evaluate the global hemodynamic response \cite{scarsoglio,CMBBE_2016,anselmino}. Multiscale modelling combines an appropriate alternation of submodels of different dimensions (e.g., zero- and one-dimensional), leading to an optimal compromise among computational efforts, required data and accuracy of the results (see, among many others, \cite{alastruey,blanco,reymond2,Mynard}). However, to the best of our knowledge, multiscale models have not fully been exploited to date to understand, from a mechanical point of view, the local arterial hemodynamics during AF.

The rationale of the work is to characterize the variability of the arterial hemodynamics during AF. This topic has been poorly investigated so far, both from computational and clinical points of view. As a consequence, the underlying mechanisms remain largely unknown: to date there is no clear evidence on how AF modifies waveform and propagation of the hemodynamic signals. For example, little is known about the propagation of the RR beating irregularity along the arterial tree and whether this cardiac disorder can significantly alter the vascular perfusion at the tissue and organ level. Only recently, the higher beat-to-beat blood pressure variability in AF has been associated to a possible vascular dysfunction, which could contribute itself to the increased cardiovascular morbidity and mortality during AF \cite{Olbers}. An explorative analysis of the hemodynamic variability at the arterial level is therefore important, since it can offer a modeling perspective for the behaviour of the arterial hemodynamics during AF, which is mostly unknown. This work represents a first approach towards an unexplored area and can be the starting point for the design of future clinical studies to investigate in vivo consequences of the pathophysiological findings emerged.


In the present work, we adopt the multiscale model proposed and validated by Guala and coauthors \cite{guala1,guala2,guala3}, which includes a lumped parameterization of the cardiac and distal circulation together with one-dimensional modelling of large-to-medium systemic arteries, to investigate in detail the local effect of AF on the arterial fluid dynamics. AF and normal sinus rhythm (NSR) conditions are simulated extracting 2000 $\mathrm{RR}$ heartbeats for each configuration to guarantee the statistical stationarity of the outcomes. To facilitate the comparison and elucidate AF-induced alterations, NSR and AF are simulated at the same mean heart rate (HR = 75 bpm). Among the fibrillated $\mathrm{RR}$ features, we therefore focus on the increased variability and decreased temporal correlation, rather than on the accelerated heart rate.

\noindent The multiscale approach allows us to estimate the main cardiac parameters, such as ejection fraction, cardiac output, stroke work, as well as to capture their variations during AF with respect to NSR. However, the study mostly involves the arterial vasculature and, more specifically, the propagation of the hemodynamic signals along the arterial tree during AF. In fact, we recall that the main aim is to quantify, through an accurate statistical analysis of pressure and flow rate waveforms, how AF affects the arterial circulation when moving away from the heart. Results will be presented to highlight the ability of the arterial system to take in the disturbances imposed at the cardiac level, whether by amplifying or damping them. In particular, pressure signals will be analyzed along the aortic trunk and brachial sections, while flow rate time series are considered in the gastrointestinal region.


\noindent The paper is organized as follows. The stochastic modelling approach is described in the Methods section, together with the parameter definition as well as the $\mathrm{RR}$ extraction. Results are proposed and discussed in the following sections (Results and Discussion), by dividing the cardiac parameters from the arterial and propagation response. Limiting aspects are reported before drawing the conclusive remarks of the study.

\section{Methods}
\label{sec:2}
\subsection{Mathematical model}
\label{sec:2.1}
\begin{figure}
\includegraphics[width=\columnwidth]{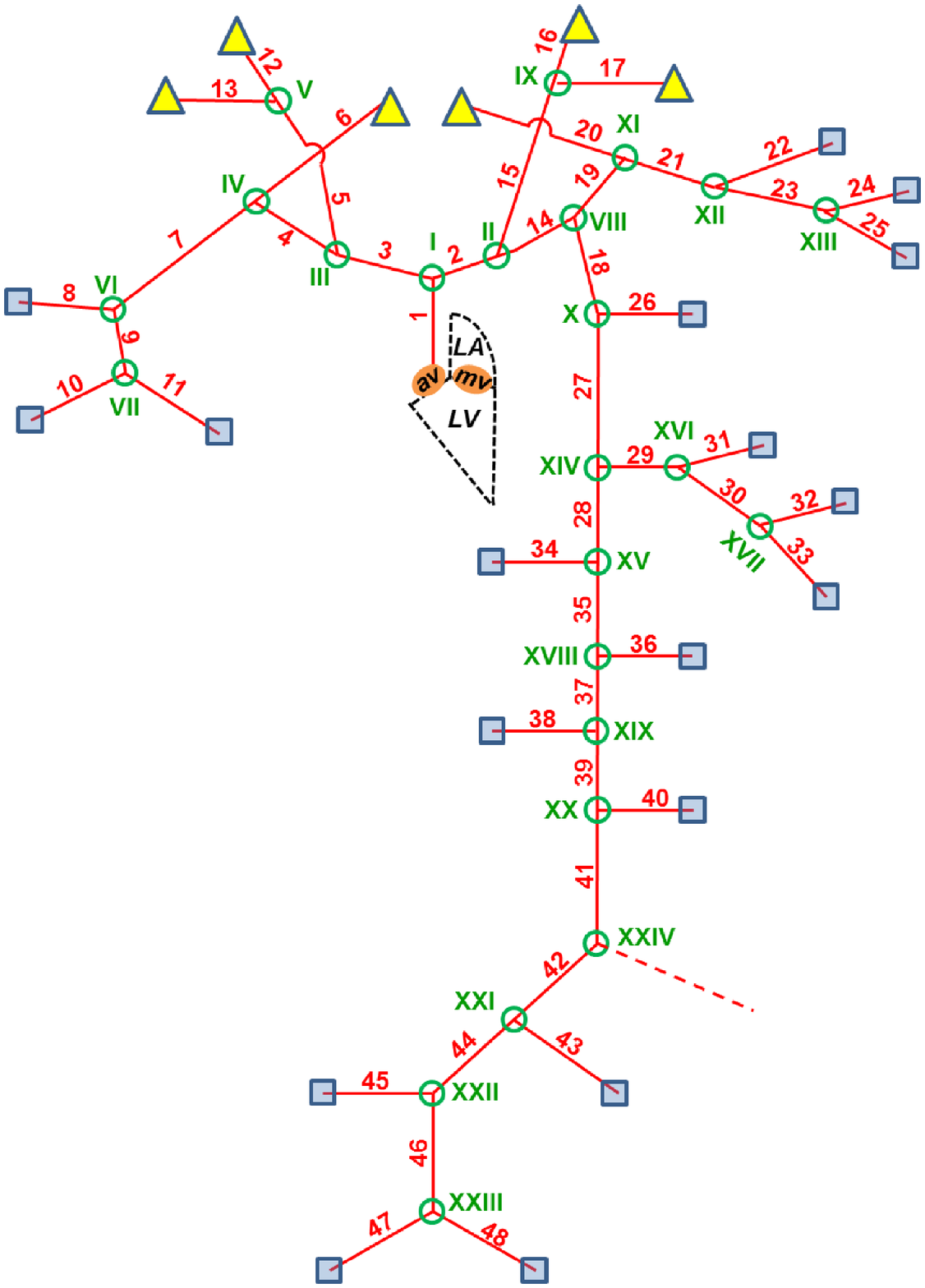}
\caption{}
\label{fig:1}
\vspace{+1cm}
\end{figure}

The present work adopts the physically-based multiscale modelling approach of the left heart and the arterial systemic circulation proposed by Guala et al. \cite{guala1}. The geometrical domain (see Fig. \ref{fig:1}) includes left heart, with mitral and aortic valves, 48 large and medium arteries, 18 micro-circulation groups and 24 arterial bifurcations. Right heart, venous return, and cerebral circulation are not modelled, while coronary circulation is not analyzed here. Both left and right arm vessels are taken into account because of their asymmetric geometry, but only the right leg is introduced exploiting the symmetry in leg arteries. The essential modelling description is briefly recalled below (all details can be found in \cite{guala1}).

Concerning large and medium arteries, vessel geometry and flow field are taken axisymmetric. Flow is laminar, being the mean Reynolds number, $Re = D_0\overline{u}$/$\nu$ ($D_0$ is the arterial diameter at the reference pressure, $P_0$=100 mmHg, $\overline{u}$ is the bulk velocity along the axis of the chosen arterial section and $\nu$ is the blood kinematic viscosity), in the range 1000$\div$100 from the central to the distal arteries. 
Arterial walls are set longitudinally tethered, producing small and only radial motions. Blood is homogeneous and Newtonian, with density $\rho$=1050 kg/$\text{m}^\text{3}$ and kinematic viscosity $\nu$=3.8 $\text{mm}^\text{2}$/s. Large and medium vessels are described through the 1D form of the mass and momentum balance equations, integrated over the transversal section
\begin{equation}
\frac{\partial A}{\partial t} + \frac{\partial Q}{\partial x} = 0,\label{eq1}
\end{equation}
\begin{equation}
\centering
\frac{\partial Q}{\partial t} + \frac{\partial}{\partial x}\int_0^R {2 \pi {u}^2 r dr} = -\frac{A}{\rho}\frac{\partial P}{\partial x} + 2\pi\nu\left
[r\frac{\partial u}{\partial r}\right]_{r=R},\label{eq2}
\end{equation}
where $x$ identifies the position along the vessel axis, $r$ is the radial coordinate, $t$ is time, $A(x,t)$ is the cross-sectional area, $P(x,t)$ is the radially-constant pressure, $Q(x,t)=\text{2}\pi \int_\text{0}^R{u r dr}$ is the flow rate, $u(x,r,t)$ is the velocity profile, and $R(x,t)$ is the vessel radius. Convective and viscous terms in the Eq. (\ref{eq2}) require to define a velocity profile, which is modelled according to \cite{guala1}
\begin{eqnarray}
 u(x,r,t)=
\begin{cases}
\tilde{u}(x,t) \,\,\, & \mbox {if } 0<r<R(x,t)-\delta(x,t),\\
\frac{{R^2(x,t)}-{r}^2}{2R(x,t)\delta(x,t)-{\delta ^2(x,t)}} \tilde{u}(x,t) \,\,\, & \mbox{if } r\ge R(x,t)-\delta(x,t).
\end{cases}
\label{eq3}
\end{eqnarray}
The flat velocity profile, $\tilde{u}$, related to the arterial flow rate, $Q$, and radius, $R$, is completed by a parabolic boundary layer of thickness $\delta$ to take into account the viscous effects. The thickness, $\delta$, is evaluated through the Womersley number, $\alpha = R\sqrt{\omega/\nu}$, where $\omega$=2$\pi$/$\mathrm{RR}$ is the cardiac pulsation, while $\mathrm{RR}$ is the cardiac beating period. 
By imposing $\alpha$=1, one has $\delta$=$\sqrt{\nu/\omega}$, that adds up to around 0.7 mm for $\mathrm{RR}$=0.8 s. By substituting the profile (\ref{eq3}) into the Eq. (\ref{eq2}), we obtain
\begin{equation}
\frac{\partial Q}{\partial t} + \frac{\partial}{\partial x} \left(\beta \frac{{Q}^{2}}{A}\right) = -\frac{A}{\rho}\frac{\partial P}{\partial x} + {N}_{4}, \label{eq5}
\end{equation}
where $\frac{\partial}{\partial x} \left(\beta \frac{{Q}^{\text{2}}}{A}\right)$ and ${N}_{\text{4}}$ express the convective and viscous terms, respectively \cite{guala1}. To solve the system (\ref{eq1})-(\ref{eq5}), a constitutive relation is needed, which links $P$ to $A$ and derives from the anisotropic non-linear visco-elastic behaviour of the arterial walls. $P$ can be therefore split into an elastic pressure component, $P_e$, and a viscous pressure component, $P_v$, as
\begin{equation}
P(x,t) = P_e(x,t) +P_v(x,t). \label{eq6}
\end{equation}
For large/medium vessels, we impose impermeability and no slip at the walls as boundary conditions.

Action of the left ventricle is described through a time-varying elastance model \cite{korakianitis,scarsoglio}, in which the elastance function, $E_{LV}(t)$, relates the left-ventricular pressure, $P_{LV}(t)$, to the left-ventricular volume, $V_{LV}(t)-V_0$ ($V_0$ is the unstressed left-ventricular volume), as follows
\begin{equation}
E_{LV}(t) = \frac{P_{LV}(t)}{V_{LV}(t)-V_0}. \label{eq9}
\end{equation}

\noindent According to the phases of the cardiac cycle, $E_{LV}(t)$ varies between its maximum ($E_{LV,max}$) and minimum ($E_{LV,min}$) values. Mitral valve is simulated as an ideal diode with a resistive term, $R_{MI}$, turning from 3 mmHg s/l to infinite when the valve closes \cite{reymond1}. The blood flow rate through mitral valve, $Q_{MI}$, results
\begin{equation}
Q_{MI}(t) = \frac{P_{LA}-P_{LV}(t)}{R_{MI}}, \label{eq12}
\end{equation}
where $P_{LA}$, the left atrial pressure, is taken constant with time.

\noindent Motion of aortic valve is reproduced through the pressure-flow rate relation as in \cite{blanco}
\begin{equation}
L \frac{dQ_{AA}}{dt} + RQ_{AA} +B|Q_{AA}|Q_{AA} = \Theta \left(P_{LV}-P_{AA}\right). \label{eq13}
\end{equation}
Here, $P_{AA}$ and $Q_{AA}$ are the pressure and flow rate at the entrance of the aorta, respectively. $L$ is the fluid inertance, $R$ represents the viscous resistance, $B$ is known as the turbulent flow separation coefficient and $\Theta$ is a function of the aortic valve opening angle, $\theta$
\begin{equation}
\Theta = \frac{(1 - \cos(\theta(t)))^4}{(1 - \cos(\theta_{max}))^4}. \label{eq14}
\end{equation}
$\theta$ varies between a minimum, $\theta_{min}$=$\ang{0}$, and a maximum, $\theta_{max}$=$\ang{75}$. The valve mechanisms are accurately modeled since the governing equation for $\theta$ accounts for the pressure difference across the valve (i.e., $P_{AA}$ and $P_{LV}$), as well as other blood flow effects such as frictional effects from neighboring tissue resistance, the dynamic effect of the blood acting on the valve leaflet, and the action of the vortex downstream of the valve \cite{korakianitis}.

The distal boundary conditions account for: (i) the effects of the missing cerebral circulation on the remaining part of the cardiovascular system at the terminal carotid and vertebral arteries; (ii) the peripheral micro-circulation groups. Both boundary conditions are expressed as three-element Windkessel sub-models \cite{guala1}, as follows
\begin{equation}
\frac{dQ_b}{dt} - \frac{1}{R_1}\frac{dP_b}{dt} = \frac{P_b-P_{ven}}{R_1 R_2 C} - \left(1+\frac{R_1}{R_2}\right) \frac{Q_b}{R_1 C}. \label{eq16}
\end{equation}
$Q_b$ and $P_b$ are the flow rate and pressure at the distal boundary region, $R_1$ represents the viscous effects for the last large/medium vessel before the distal group, and $R_2$ and $C$ stand for the viscous and elastic reactions from the distal area, respectively. $P_{ven}$=5 mmHg is the constant pressure of the micro-circulation venous region.

We set the conservation of the total pressure and mass at each arterial junction. If the generic parent artery 0 divides into the daughter arteries 1 and 2, at the bifurcation we have
\begin{equation}
\begin{cases} P_0(L_0,t) + \frac{1}{2}\rho\overline{u}_0^2(L_0,t) = P_1(0,t) + \frac{1}{2}\rho \overline{u}_1^2(0,t), \\ P_0(L_0,t) + \frac{1}{2}\rho \overline{u}_0^2(L_0,t) = P_2(0,t) + \frac{1}{2}\rho \overline{u}_2^2(0,t), \\ Q_0(L_0,t) = Q_1(0,t) + Q_2(0,t),
\end{cases} \label{eq17}
\end{equation}
where $L$ is the vessel length and $\overline{u}$ is the mean blood velocity.

The mathematical model is solved numerically through a Runge-Kutta Dis\-con\-tin\-u\-ous-Galerkin method. First, space is discretized by a Discontinuous-Galerkin approach, then time evolution is solved by an explicit second order Runge-Kutta scheme. Simulations are carried out with a time step equal to 10$^{\text{-\text{4}}}$ s and a minimum element length of 2.5 cm. Solutions are not dependant on initial conditions (constant pressure equal to 100 mmHg and no flow everywhere) and convergence is reached after about 7 heart cycles.

Setting of the geometrical and hemodynamic parameters is performed according to previous studies. In particular, sizes and distal values of the arterial vasculature are those reported in \cite{reymond2}, while the arterial wall thicknesses and viscoelasticity coefficients are those indicated by \cite{blanco}. Apart from the minimum ($E_{LV,min}=0.08$ mmHg/ml) and maximum ($E_{LV,max}=2.88$ mmHg/ml) values of the time-varying elastance, $E_{LV}(t)$, as well as $V_0=14.00$ ml and $P_{LA}=8.33$ mmHg, all the other cardiovascular parameters are set as in \cite{guala1}, in order to reproduce the physiological behaviour of the heart-systemic circulation, for an healthy young man with normal beating (HR= 75 bpm).

\subsection{Beating features: normal sinus rhythm and atrial fibrillation}
\label{sec:2.4}

In order to identify the effects of AF on the arterial response, we force the cardiovascular system with two different sequences of cardiac interbeats, $\mathrm{RR}_{NSR}$ and $\mathrm{RR}_{AF}$.

\noindent NSR beating is a remarkable example of pink noise \cite{hayano}, thus normal sinus intervals, $\mathrm{RR}_{NSR}$, are extracted from a pink-correlated Gaussian distribution (mean value $\mu=$0.8 s, standard deviation $\sigma=$ 0.06 s), as usually observed in sinus rhythm \cite{hennig,scarsoglio}. Differently to the white noise which is uncorrelated, pink noise introduces a temporal correlation, which is a common feature of the NSR beating.

\noindent The AF distribution is fully described \cite{hennig,hayano} by the  superposition  of  two   statistically  independent  time intervals,  $\mathrm{RR}_{AF}=\varphi+\eta$. Time intervals $\varphi$ are  obtained  from  a  correlated  pink  Gaussian  distribution,   $\eta$  are instead  drawn  from  an uncorrelated exponential distribution (rate parameter $\gamma$). The resulting AF intervals, $\mathrm{RR}_{AF}$, are thus extracted from an uncorrelated Exponential Modified Gaussian distribution (mean value $\mu=$ 0.8 s, standard deviation $\sigma=$ 0.19 s, rate parameter $\gamma=$ 7.24 [Hz]), which is the most common (60-65\% of the cases) unimodal distribution during AF \cite{hennig,hayano}. The AF beating results less correlated and with a higher variability than NSR, as clinically observed \cite{hennig,Bootsma,hayano}.

\noindent The $\mathrm{RR}$ parameters in NSR and AF conditions are suggested  by the available literature \cite{hennig,Sosnowski,hayano}, and by considering that the coefficient of variation, $cv$, is around 0.24 during AF \cite{Tateno}. Once these $\mathrm{RR}$ intervals have been validated and tested over clinically measured beating \cite{hennig,hayano}, we adopt them as the most suitable and reliable $\mathrm{RR}$ sequences to mimic NSR and AF. More details on the beating features are offered elsewhere \cite{scarsoglio,anselmino}. To facilitate the comparison, both series are taken at the same mean HR (75 bpm). Leaving aside the accelerated beating, the focus is thus on the other two AF beating characteristics: a reduced temporal correlation and an increased temporal variability ($\sigma_{AF}>\sigma_{NSR}$) with respect to the NSR.

The cardiac force-frequency relationship, which is especially important in AF for the $\mathrm{RR}$ variability, is here taken into account since the elastance model adopted is defined through an activation function, which explicitly depends on the $\mathrm{RR}$ length \cite{korakianitis}. As for the contractile strength, in patients with chronic AF and concomitant cardiomyopathy, a ventricular contractile dysfunction may develop \cite{Tanabe}, translating into a reduction of the maximum LV elastance, $E_{LV,max}$. However, a possible dysfunction solely induced by AF is far from being clear and contrasting trends emerge in literature \cite{Tanaka}. Given the lack of definitive data \cite{scarsoglio} during AF, the same left ventricular contractility (i.e., $E_{LV,max}$) was assumed in both NSR and AF.

\begin{figure}
\includegraphics[width=\columnwidth]{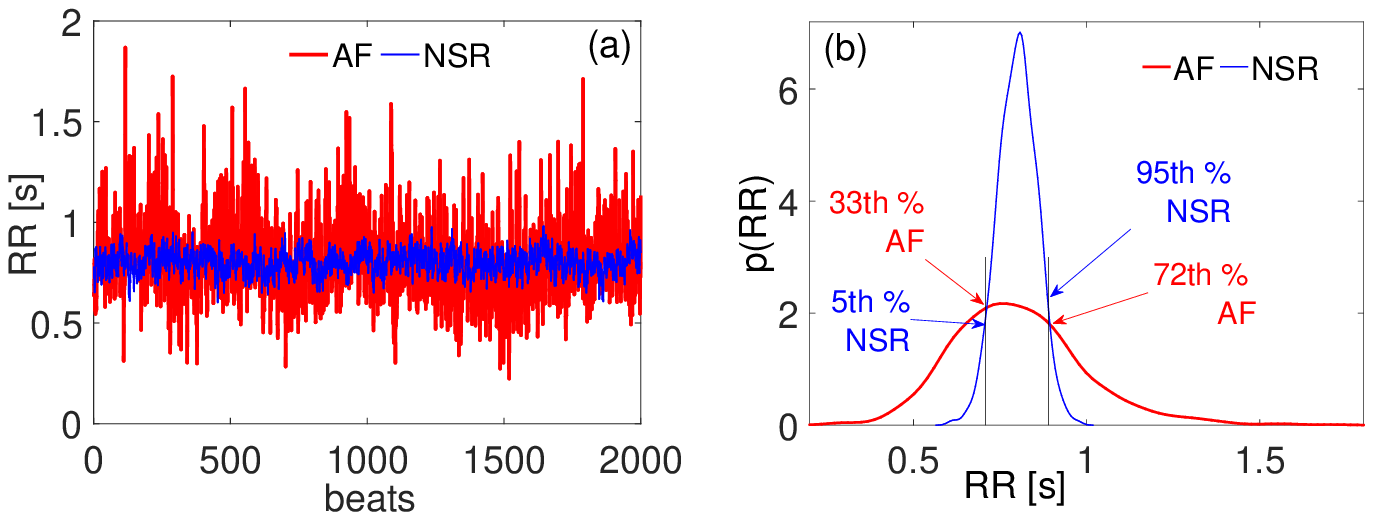}
\caption{}
\label{fig:2}
\vspace{+1cm}
\end{figure}

To accurately model NSR and AF beating features, 2000 heartbeats are extracted for each configuration to guarantee the statistical stationarity of the results. The $\mathrm{RR}_{NSR}$ and $\mathrm{RR}_{AF}$ series, corresponding to about half an hour of cardiac activity, together with the corresponding probability density functions (PDFs) are reported in Fig. \ref{fig:2}. The main statistics of the $\mathrm{RR}$ intervals are compared in Table \ref{tab:1}.

\section{Results}
\label{sec:3}

Through the modelling approach, we gain the complete hemodynamic signals for the left-ventricular pressure and volume, as well as for the pressure and flow rate at each arterial section. We identify some quantities of interest concerning the left ventricle and the vessels, and calculate their PDFs from the 2000 corresponding data. Each PDF is characterized through the related mean value, $\mu$, standard deviation, $\sigma$, and coefficient of variation, $cv=\sigma/\mu$. To highlight AF-induced changes, for each hemodynamic parameter considered we carry out the analysis of percentile variation between NSR and AF: firstly, we define the 5th and 95th percentiles in NSR as reference thresholds; secondly, during AF we evaluate to which percentile each of the NSR thresholds corresponds. In so doing, independently from the PDF shape, we can quantify whether AF is able to enhance the probability of reaching extreme values, either uncommonly high or low. An example is reported in Fig. \ref{fig:2}b, for the $\mathrm{RR}$ distribution (and analogously applies for all the hemodynamic variables considered in the following). The $\mathrm{RR}$ values individuated by the $5th$ and $95th$ percentiles in NSR correspond to the $33rd$ and $72nd$ percentiles in AF, respectively.

\noindent We present the main results for both left ventricle and arteries in the three following subsections. The first one involves the main cardiac and hemodynamic parameters, the second one regards arterial pressure patterns along the systemic tree, and the last one inquires into the mechanisms of wave propagation and reflection.

\subsection{Cardiac and hemodynamic parameters}
\label{sec:3.1}

The most important cardiac and hemodynamic parameters are analyzed in terms of mean, standard deviation and $cv$ values, during NSR and AF, as reported in Table \ref{tab:2}. We define the end-diastole as the time immediately before the beginning of the isovolumic contraction, when the mitral valve closes. At this point we compute the end-diastolic left ventricular pressure, $EDP$, and end-diastolic left ventricular volume, $EDV$. The end-systole is taken as the instant at which aortic valve closes, and here the end-systolic left-ventricular pressure, $ESP$, and volume, $ESV$, are evaluated. We also compute the following left-ventricle cardiac parameters: the stroke volume, $SV = EDV - ESV$ [ml], the ejection fraction, $EF = SV/EDV \cdot 100$ [\%], the cardiac output $CO = SV \cdot HR$ [l/min], and the stroke work, $SW$ [J], as the area of the PV loop. Available literature data are also compared with the present results, by adopting the following notation for the average variations during AF with respect to NSR: $+$ increase, $−-$ decrease, $=$  no substantial variation, $/$ no data available.

Starting with the NSR condition, we find that the present average values, $SV = 66.57$ ml, $EF = 55.38$ \%, $CO = 4.99$ l, $SW = 0.92$ J, are in good agreement with the physiological values \cite{guyton,westerhof1,Klabunde}. The aortic systolic ($P_{AA,syst}= 120.50$ mmHg) and diastolic  ($P_{AA,dia}= 71.00$ mmHg) pressures also fall within the physiological aortic pressure range \cite{guyton,caro,Klabunde}.

In terms of AF variations with respect to NSR (last two columns of Table \ref{tab:2}), a qualitative good correspondence is observed between present outcomes and literature results. Indeed, almost all trends here found agree with clinical data. To test the statistical significance of the mean variations between NSR and AF, we performed a t-test for all the couples (NSR and AF) of the hemodynamic variables, using the hypothesis of unequal variances (Welch's t-test). p-values are reported in the fifth column of Table 2. Apart from the systolic pressure, $P_{AA,sys}$, all the hemodynamic variables show a significant difference between NSR and AF.

\noindent A more detailed quantitative assessment involving the specific values assumed by the hemodynamic variables is out of the scope of the present work, as we recall that: (i) clinical data are heterogeneous differing in age, sex, type of AF (e.g., chronic, paroxysmal, ...), concomitant pathologies, AF treatment (e.g., cardioversion, catheter ablation, drugs  treatment, ...), and follow-up periods. This heterogeneity has no counterpart in the modelling approach, which is exploited to analyze the global heart-arterial response to AF for a generic young healthy patient. (ii) To facilitate the comparison between NSR and AF, the two configurations are here simulated at the same mean HR (75 bpm). As a consequence, percentage variations of mean values are small and, in general, underestimated. Keeping these two aspects in mind, it is worth noting that, given the same HR, the irregular and uncorrelated AF beating features are able themselves to promote hemodynamic variations, which go in the same direction as clinically observed.

\begin{figure}
\includegraphics[width=\columnwidth]{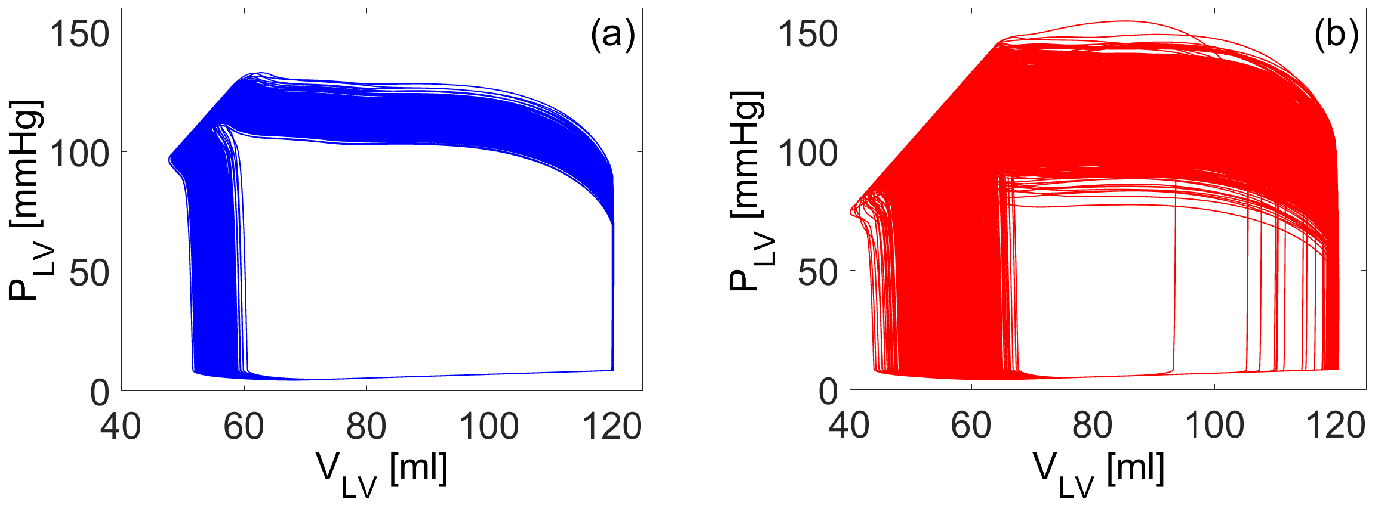}
\caption{}
\label{fig:3}
\vspace{+1cm}
\end{figure}

\noindent Despite end-diastole volume and pressure are the variables which are mostly affected by the assumption of a constant left-atrial pressure, the PV loops reported in Fig. \ref{fig:3}a correctly describe the characteristic phases of the cardiac cycle during NSR \cite{guyton}, while the PV loops in AF (Fig. \ref{fig:3}b) are fully representative of the much higher variability induced by AF at the cardiac level (about 3 times higher than NSR).

\subsection{Arterial pressure patterns}
\label{sec:3.2}

\begin{figure}
\centering
\includegraphics[width=\columnwidth]{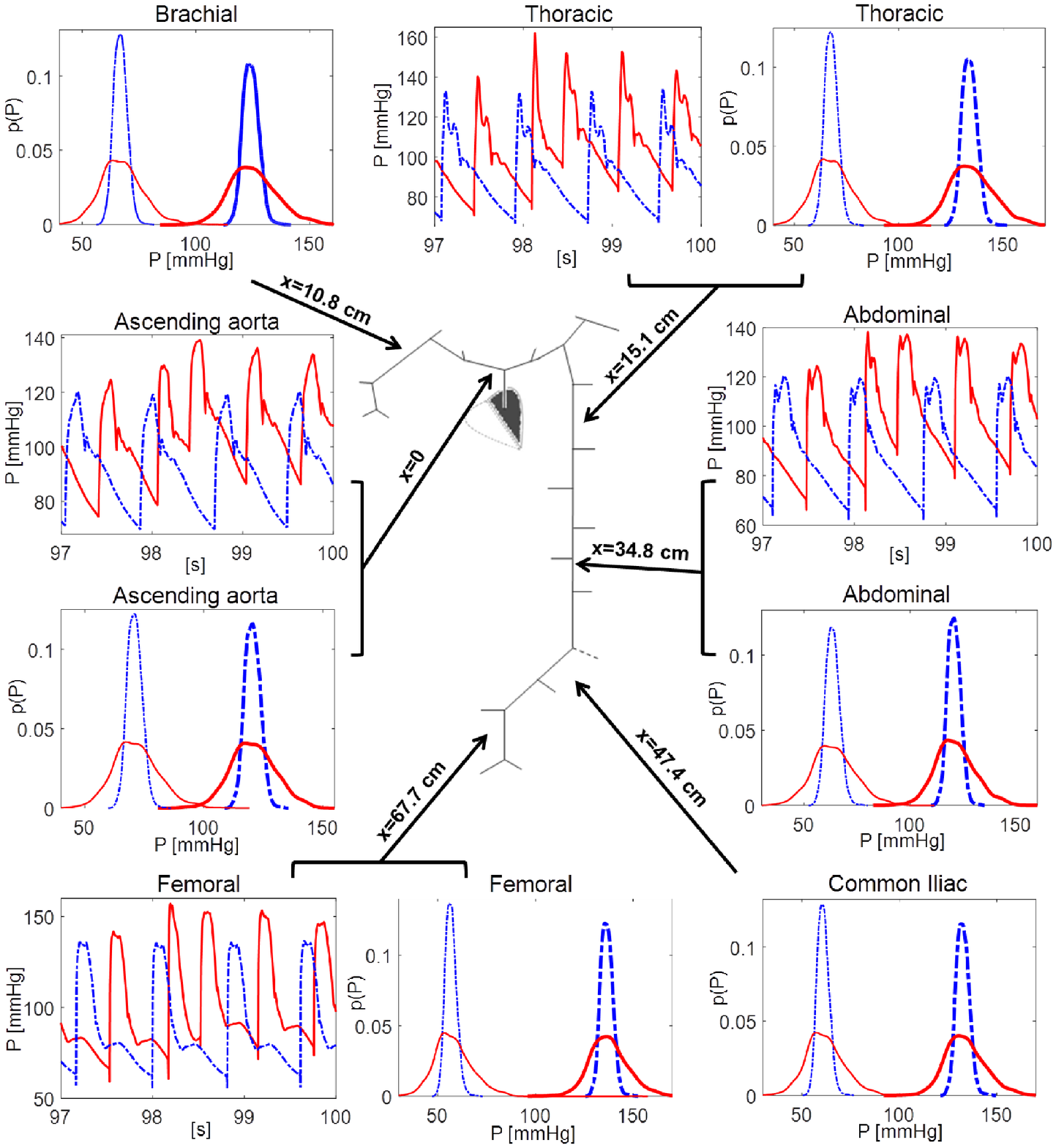}
\caption{}
\label{fig:4}
\vspace{+1cm}
\end{figure}

To analyze the arterial patterns, we introduce the distance from the heart, $x$ [cm] ($x=0$ is the beginning of the ascending aorta), which is purely representative of the arterial location and does not account for the different number of bifurcations that might have occurred in the path from heart to the location $x$. Regarding large/medium arteries, we consider the systolic, diastolic and pulse pressures. PDFs of systolic and diastolic pressures are reported in Fig. \ref{fig:4}, together with a portion of the pressure time-series at some representative arterial regions moving away from the heart at different $x$. It is evident that systolic and diastolic pressure values cover wider intervals during AF than in NSR, as a consequence of the greater variability of the local pressure values. Moreover, the AF-induced variability extends to the whole arterial system, including the aortic trunk as well as lower and upper limbs.

\begin{figure}
\includegraphics[width=\columnwidth]{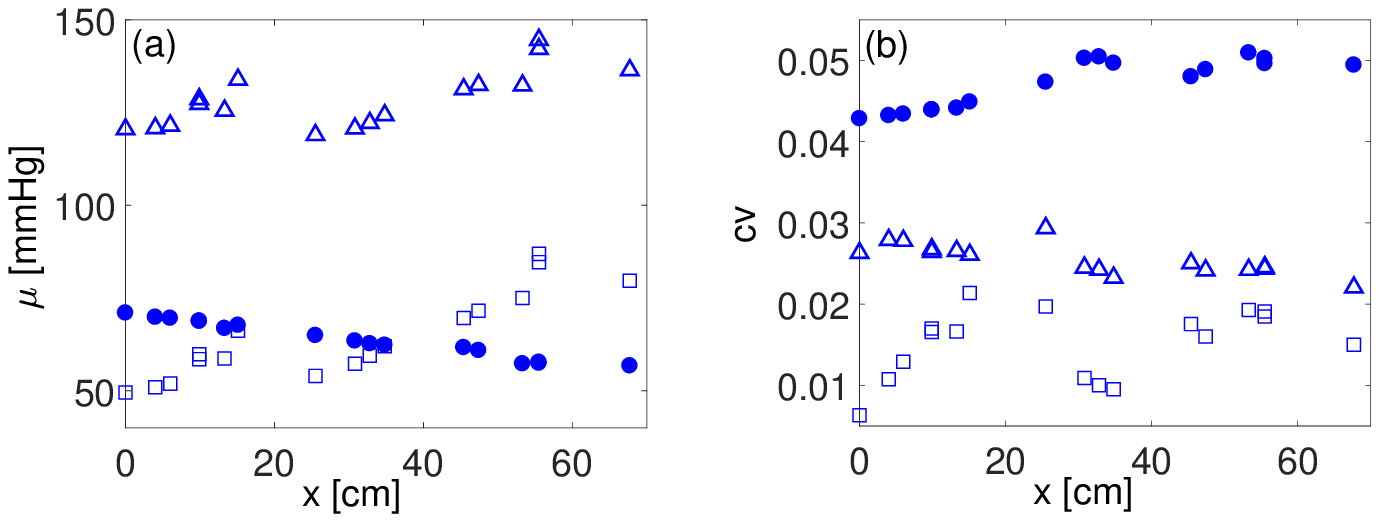}
\caption{}
\label{fig:5}
\vspace{+1cm}
\end{figure}

\noindent As shown in Fig. \ref{fig:5}a, in NSR, mean systolic pressures grow along the aorta, while diastolic pressures reduce, resulting in an increase of the pulse pressure, $PP$, thereby confirming the normal physiological trend \cite{Klabunde,Vosse}. NSR oscillations around mean values are in the intervals 2-3\%, 4-5\% and 1-2\% for systolic, diastolic and pulse pressures, respectively. Fig. \ref{fig:5}b indicates that $cv$ values, in NSR, grow for diastolic pressures and decrease for systolic ones, moving away from the heart. This means that the arterial system tends to amplify diastolic pressure oscillations, reducing the systolic ones during NSR. However, the trends are in general non-monotonic and, especially for the pulse pressure, quite variable along the aorta, already revealing in physiological conditions the complexity of the arterial fluid dynamics.

\begin{figure}
\includegraphics[width=\columnwidth]{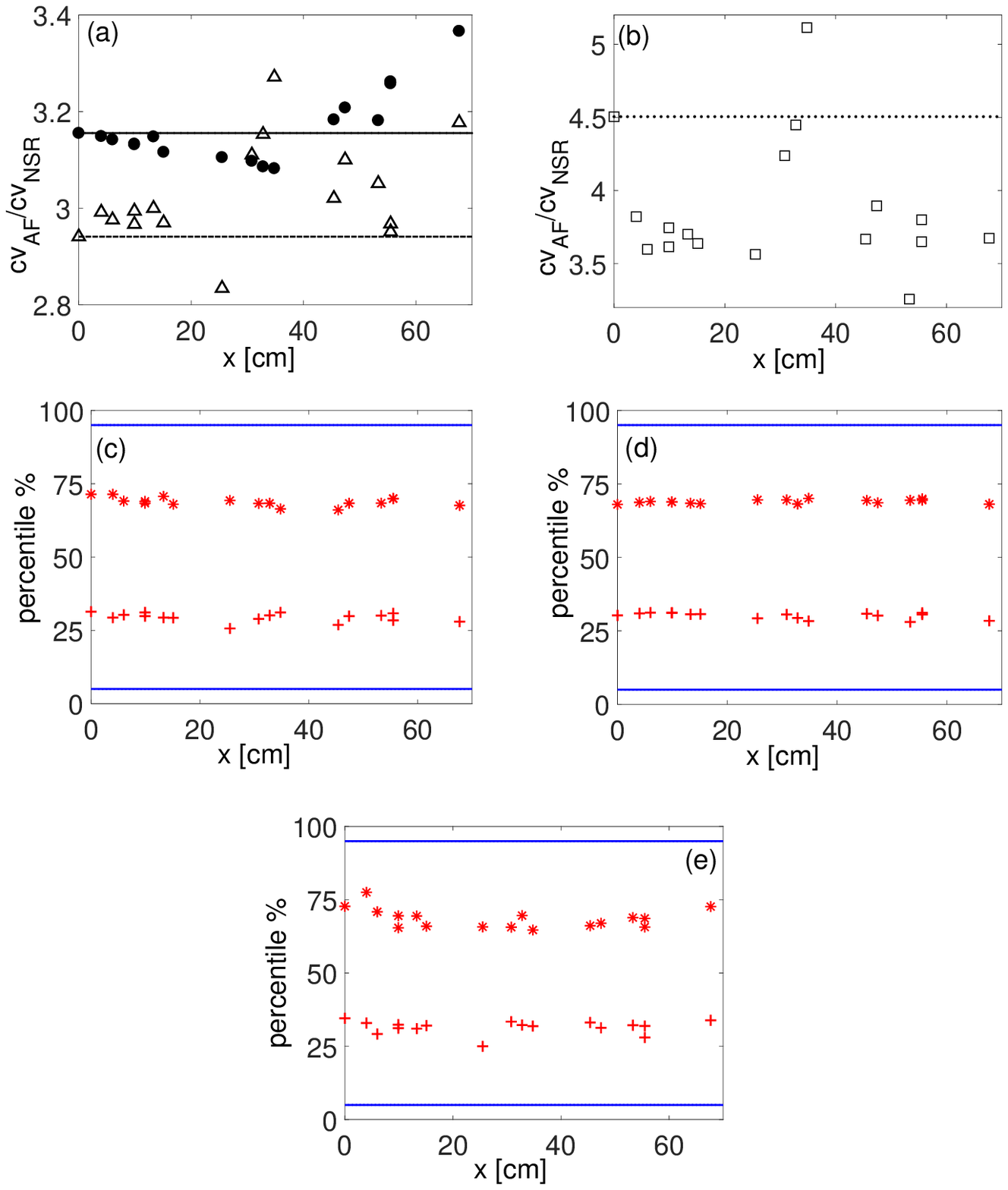}
\caption{}
\label{fig:6}
\vspace{+1cm}
\end{figure}

During AF, mean values of the systolic, diastolic and pulse pressures vary minimally site by site. On the contrary, the intervals of fluctuations around mean values in AF increase, becoming 7-8\%, 14-19\% and 3-10\% for systolic, diastolic and pulse pressures, respectively. These results are evident from the PDFs for systolic and diastolic pressures shown in Fig. \ref{fig:4}, but are better highlighted by the $cv$ ratios between AF and NSR along the arterial tree, as depicted in Fig. \ref{fig:6}a,b. The $cv$ ratios vary between 3 to 4, implying a much higher variability during AF with respect to NSR. Moreover, Fig. \ref{fig:6}a,b indicate the ability of the arterial system to amplify or damp the introduced variability with respect to the ascending aorta. In fact, the constant horizontal lines in panels a and b represent the $cv$ ratio values at the beginning of the aorta (3.15 for the diastolic, 2.95 for the systolic, 4.5 for the pulse pressures). If data along the $x$ direction fall above the horizontal line, an amplification of the variability occurs, while if they fall below the horizontal line, the arterial system is able to damp the variability introduced at the aortic root level. In general, $cv$ ratios of both systolic and diastolic pressures do not display a monotonic trend along the arterial pathway (Fig. \ref{fig:6}a), meaning that the response is not solely influenced by the distance from the heart, $x$ [cm]. However, it can be noted that, for $x>30$ cm (that is, overcoming the abdominal region), both systolic and diastolic $cv$ ratios tend to be amplified. Apart from a few exceptions near the abdominal zones (around $x=30-35$ cm), the $cv$ ratio of the pulse pressure along $x$ (Fig. \ref{fig:6}b) is instead kept well below ($cv$ between 3.5 and 4) the starting value at the beginning of the aorta ($cv$=4.5, evidenced by the horizontal dashed line).

Given the $5th$ and $95th$ reference percentiles of the PDFs for systolic, diastolic and pulse pressures in NSR, Fig. \ref{fig:6}c,d and e represent how these percentiles modify in AF along the arterial tree. Pressure percentiles are relatively constant, with no clear trend to increase or decrease along $x$, and with a greater variability for the pulse pressure (Fig. \ref{fig:6}e). The lateral areas of the PDFs grow in AF with respect to NSR, strongly increasing the probability of extreme values. In all the cases reported, the $5th$ and $95th$ percentiles in NSR correspond in AF to about the $30th$ and $70th$ percentiles, respectively.

\subsection{Wave propagation and reflection}
\label{sec:3.3}

Since in NSR wave propagation and reflection along the arterial tree is a balanced interplay between forward and backward signals, we inquire here whether AF is able to perturb this equilibrium. Therefore, for both NSR and AF, at different arterial points we evaluate the phase velocity, $c$, through the foot-to-foot method \cite{westerhof1}, and the Reflection Magnitude \cite{murgo,westerhof2,li,westerhof1}, $RM$. Chosen a vessel section, beat by beat, $c$ is an estimate of the local wave speed within the high frequency range, while $RM$ is the ratio between the amplitudes of the local backward and forward pressure signals.

\begin{figure}
\includegraphics[width=\columnwidth]{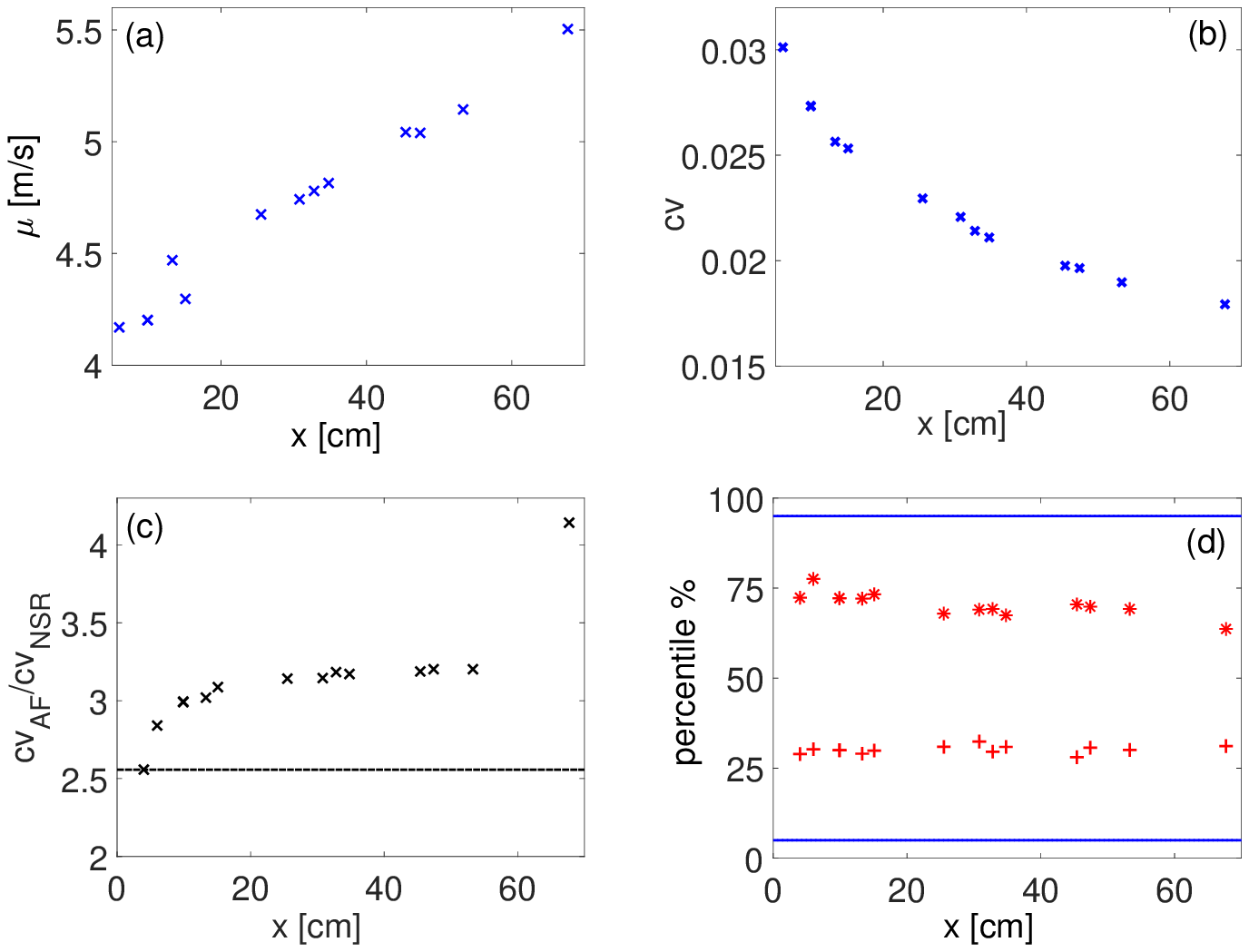}
\caption{}
\label{fig:7}
\vspace{+1cm}
\end{figure}

Fig. \ref{fig:7} shows different aspects of the behaviour of the phase velocity, $c$, in NSR and AF. Mean and $cv$ values for $c$ along the arterial pathway in NSR are shown in panels a and b, respectively. Mean phase velocity rises with the distance from the heart, $x$, within the interval 4.2-6.5 m/s, in agreement with what observed in literature \cite{caro,Klabunde}. Fluctuations around mean values regularly and smoothly decrease towards the distal regions, going from 3.5\% to 1.5\% moving away from the heart (Fig. \ref{fig:7}b). This means that, even if the signal maintains the same variability $\sigma$, the phase velocity fluctuations are less impacting since the mean velocity increases.

\noindent Regarding AF, as already observed for the pressures, mean values of $c$ (not shown here) practically do not change as imposing $\mathrm{RR}_{AF}$, in reference to the results for $\mathrm{RR}_{NSR}$ (percentage differences are below 1\%). The average trends in NSR and AF along $x$ are pretty similar, with small differences at more distal zones. Instead, fluctuations around the mean values of $c$ largely upsurge in AF with respect to the NSR, as one can see from Fig. \ref{fig:7}c. The $cv$ ratio between AF and NSR of $c$ not only is always greater than 1, but for all the arterial locations, $x$, is also higher than the value individuated at the beginning of the ascending aorta, where the $cv$ ratio is about 2.5 (as evidenced by the constant horizontal line). This implies that for increasing distances $x$, during AF the phase velocity fluctuations (in terms of $cv$) are not damped, contrarily to what happens in NSR. Fig. \ref{fig:7}d indicates, for different distances $x$, to which AF percentiles the $5th$ and $95th$ NSR percentiles correspond. As for the pressure PDFs, the phase velocity PDFs enlarge dramatically in AF, too. In all the arterial regions considered, the $5th$ percentiles in NSR become around $30th$ in AF, while the $95th$ percentiles in NSR are in the interval $60th$-$80th$ in AF, with a decreasing worsening trend towards the peripheral regions.

\noindent Trends of the phase velocity in AF have not been analyzed in detail so far, thus they represent a novel aspect of the present work. We recall that phase velocity is determined by blood density and distensibility \cite{caro}. Assumed the former as constant, distensibility is a property of the blood vessel, being at the same time dependent on pressure and frequency. Since during AF pressure waveforms and heart rate exhibit a higher variability, it is reasonable that the phase velocity computing, which is carried out through approximation methods based on the hemodynamic signals, inherits such a higher variability, as observed in Fig. \ref{fig:7}c and \ref{fig:7}d.

\begin{figure}
\includegraphics[width=\columnwidth]{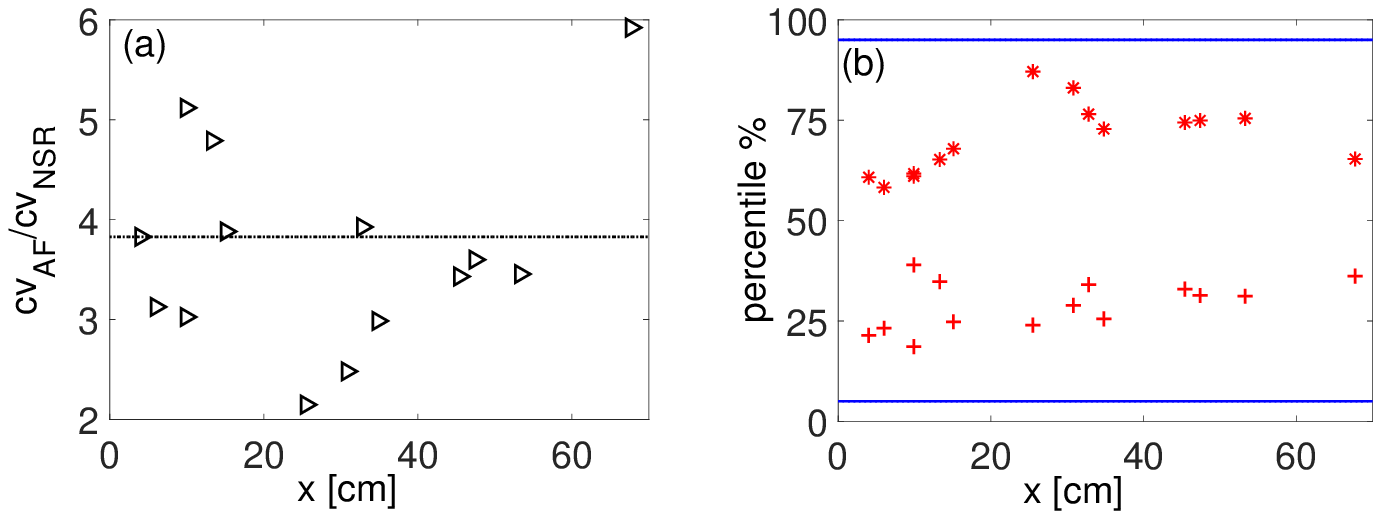}
\caption{}
\label{fig:8}
\vspace{+1cm}
\end{figure}

Concerning $RM$, oscillations around mean values are definitely much more pronounced in AF than in NSR, with a $cv$ ratio between AF and NSR in the interval 2-6 along $x$ (Fig. \ref{fig:8}a), which is general well above 1. There is an extreme variability in the magnitude of fluctuations in AF, in reference to the NSR, which is different according to the arterial section. The highest amplification in the $RM$ fluctuations during AF is estimated at the femoral artery, where also fluctuations of the phase velocity are very high (see Fig. \ref{fig:7}c). Along thoracic and abdominal aorta ($20<x<40$ cm), the $cv$ ratio of $RM$ is lower than or equal to that of the ascending aorta, that is about 4. Instead, along the brachial artery (around $x=15-20$ cm), the $cv$ ratio results higher than at the entrance to the aorta. The percentiles in AF, corresponding to the $5th$ and $95th$ percentiles in NSR, are given in Fig. \ref{fig:8}b. It is visible that the PDFs of $RM$ in AF have a higher width than in NSR, with the $5th$ percentile during NSR reaching the range between $20th$ and $40th$ in AF, and the $95th$ percentile in NSR ending up to be within $60th$ and $85th$ during AF. The longest tails in the PDFs during AF are estimated along abdominal aorta, with the $5th$ percentile becoming between $20th$-$35th$ and the $95th$ percentile reaching the interval between $70th$-$85th$. The high variability along $x$ observed for $RM$ through both the $cv$ ratio and the percentiles highlight how differently waves reflect at the arterial junctions in presence of AF.

\noindent As a conclusive remark, it should be mentioned that $x$ only accounts for the distance from the heart and not for the number of bifurcations occurred so far. This may also explain some of the irregularity in the previous plots (e.g., Fig. \ref{fig:5} and \ref{fig:8}) along $x$.

\section{Discussion}
\label{sec:4}

Results so far presented highlight several features of the arterial hemodynamics during AF that deserve to be discussed. We first focus on the behaviour of the coefficient of variation, $cv$, along the arterial tree for systolic and diastolic pressures. We find that the value of $cv$ for diastolic pressure is higher than for systolic pressure at each arterial section. Furthermore, $cv$ reduces for systolic pressures and increases for diastolic pressure with the distance from the heart, $x$. This behaviour is already present in NSR, but is not linearly magnified during AF. In fact, the ratios of $cv$ between AF and NSR (always greater than 1) are everywhere higher for diastolic pressures than for the systolic ones, with a few exceptions along abdominal aorta. In addition, these ratios are not the same site by site and tend to grow for both pressures at more distal areas. Moreover, it can be observed that: (i) diastolic pressures require more or the same number of beats than systolic pressures to overcome the effects of the perturbation, once a variable $\mathrm{RR}$ disturbance is introduced; (ii) distal sites have a greater inertia than central ones in restoring the equilibrium condition. The emerging scenario is that the arterial system is not able to completely absorb and damp the AF-induced variability. On the contrary, the introduced fluctuations tend to grow towards the peripheral regions, being the diastolic pressure the hemodynamic variable which is mostly affected by the AF variability.

A second aspect drawn from this study is that the tails of the distributions for all the quantities under exam are decisively more pronounced in case of AF than in NSR. Thus, in AF, the probability to have extremely low/high values of pressures, phase velocities and coefficients of reflection is much higher than in NSR. This is immediately visible by looking at $5th$ and $95th$ percentile variations experienced by all the quantities analyzed in AF. In particular, the $5th$ percentile tends to be around $30th$ and the $95th$ tends to be around $70th$, for almost all the variables. These results are not too different with the distance from the heart, $x$, apart from $RM$, which reveals more variable percentiles along $x$. In general, if left and right tail values are summed, the 10\% probability of extreme values in NSR becomes 60\% in AF, which is comparable to the median value. This means that an extremely low/high value which is rarely reached in NSR becomes common and frequently attained in AF.

In light of all the above, we find that AF causes non-negligible fluctuations in pressures, as well as phase velocities and magnitudes of reflection, provoking an enlargement of the corresponding PDFs, with respect to the NSR. Notwithstanding, changes in the magnitude of fluctuations
for pressures seem to vary depending on the specific arterial point, as the $cv$ ratios between AF and NSR (Fig. \ref{fig:6}a,b) demonstrate. This aspect can be explained if we have in mind that pressure and flow rate signals, produced by the contracting activity of the heart, are nothing but waves. These last ones travel at the finite and locally variable speed (estimated, beat by beat, through $c$) and are reflected, most of all, because of the arterial bifurcations (the local magnitude of reflection at each beat is given by $RM$). Therefore, chosen an arterial section, it is possible to identify 4 factors responsible for the definition of the local pressure signal: (i) the initial pressure signal at the entrance to aorta, (ii) the local phase velocity, (iii) how waves are reflected at the nearest bifurcations to the chosen site, (iv) the distance between the chosen site and the nearest sites of reflection. This work shows that the first 3 mentioned elements experience substantial modification under AF. Thus, beat by beat, we can recognize these three sources of alteration of the local pressure signal. The first one is related to the fibrillated beating sequence, $\mathrm{RR}$, which varies the initial pressure signal at the ascending aorta level. The other two factors are the mechanisms of wave propagation and reflection, which physiologically vary site by site. We observe, through the $cv$ ratios and percentile variations of $c$ and $RM$, that AF alters the local wave dynamics non-uniformly along the arterial tree. The combination of these three sources of variation is responsible of the complex hemodynamic scenario here described.

To complete the analysis we briefly discuss how mean flow rates per beat, $\overline{Q}$ [ml/beat], are affected by AF at the gastrointestinal bifurcations from the aorta present in the model (see Table \ref{tab:3}). As for the pressures, the average values of $\overline{Q}$ remain almost unvaried comparing AF and NSR, at each site. However, the higher variability induced by AF is well represented by the $cv$ ratios and percentile variations of Table \ref{tab:3}. It is worth noting that, even if the $cv$ ratios are always higher than 1, at the abdominal bifurcations the fluctuations are damped with respect the ascending aorta. As for the percentile variations, the $5th$ and $95th$ percentiles in NSR correspond to about the $25-30th$ and $65-70th$ percentiles in AF, respectively. Overall, the flow rates confirm substantially the dynamics observed for the pressure signals.

The irregular hemodynamic pattern and the repeated exposure to extreme flow rate and pressure values induced by AF could promote local suffering or dysfunction at the arterial level in the long-term. In fact, as very recently observed by Olbers et al. \cite{Olbers}, the higher beat-to-beat variability triggered by AF could be associated with an increased risk for cardiovascular morbidity and mortality, not entirely explained by thromboembolism. Hemodynamic effects of prolonged anomalous pressure levels and local blood flow impairment deserve deeper examination in the future, from both clinical and computational points of view.


\section{Limitations}
\label{sec:5}

One of the limitations of the present work deals with the fact that the model is not equipped with short-term autoregulation mechanisms. In other words, we do not consider any internal adaptive mechanism able to adjust the arterial system response to the beat duration. However, this simplifying assumption is often adopted in cardiovascular modelling \cite{blanco,korakianitis,alastruey,reymond2,Mynard,scarsoglio}. Long-term structural remodelling effects due to the chronic persistence of AF are also neglected. Other limiting aspects are related to the modelling approach. We do not consider the coronary circulation and the small portion of flow rate directed to oxygenate the myocardium and other heart components. Despite the fact that little is known about the AF effects on the cerebral hemodynamics, the absence of the cerebral circulation has no significant repercussions on the results provided trough the model. In fact, the influence of the cerebral districts on the rest of the modelling is adequately replaced by suitable boundary conditions. In the end, the assumptions of a non-contractile left atrium and a constant left atrial pressure mainly affect the left ventricle hemodynamics, by underestimating the fluctuations of the end-diastolic phase in both NSR and AF.

\section{Conclusions}
\label{sec:6}

The present approach proves to be a powerful \textit{in-silico} tool in revealing pressure alterations along the arterial tree during AF, in reference to the NSR. To focus on the AF-induced variability and facilitate the comparison with NSR, both NSR and AF conditions have been simulated at the same mean heart rate (75 bpm). Given this, the irregular and uncorrelated AF beating features are able themselves to promote hemodynamic variations, which qualitatively agree with literature data available for cardiac parameters. Before entering the arterial tree, a much higher variability (3 times higher than NSR) is induced by AF already at the ventricular level. Within the arterial tree, the introduced fluctuations persist and tend to grow in terms of pressure towards the distal regions. The diastolic pressure, in particular, is the most prone to the AF variability. Nevertheless, extreme values in NSR become rather common during AF, and AF is also able to locally alter the wave dynamics, by modifying the interplay between forward and backward signals. A similar scenario is observed for the flow rates, leading to a generalized irregularity of the local perfusion. Thus, the arterial system is not able to completely absorb the fluctuations inserted through the fibrillated beating, $\mathrm{RR}_{AF}$. For flow rates, the variability is damped with respect to the entrance, but remains still quite higher than in NSR. For pressures and phase velocity, fluctuations are amplified with respect to the ascending aorta, reaching values up to 3 to 4 times higher than NSR towards the peripheral circulation. In general, the response along the arterial tree strongly depends on the region considered. It emerges that the fluctuation trends along $x$ are not monotonic and with non-trivial features.


\noindent The proposed work represents a first attempt to characterize the arterial hemodynamics in presence of AF. Considering that heart rate is the same in both NSR and AF conditions, the sole heart rhythm variation is able itself to modify the regular and periodic hemodynamics at the arterial level. In terms of pressure and peripheral perfusion, the evidenced alteration could suggest in the long-term a dysfunction or local suffering due to repeated exposure to irregular and extreme values. All these aspects surely deserve further investigation in the future to provide new insights into the AF hemodynamics, especially in those regions - such as the cerebral and coronary circulations - where clinical evidence is still lacking.

\begin{acknowledgements}
The authors would like to thank Andrea Guala for the valuable and precious support with the model settings and numerical simulations.
\end{acknowledgements}

\section*{Compliance with ethical standards}
\textbf{Conflict of interest} The authors declare that they have no conflict of interest.

\section*{Figure Legends}

\noindent \textbf{Figure 1.} Schematic domain of the mathematical model. Large and medium arteries are indicated by continuous lines numbered through Arabic numerals, according to \cite{reymond2}. The arterial bifurcations are depicted by circles and numbered through Roman numerals. Squares and triangles represent the distal micro-circulation groups and the cerebral boundary conditions, respectively. $mv$: mitral valve, $LA$: left atrium, $LV$: left ventricle, $av$: aortic valve.

\noindent \textbf{Figure 2.} (a) Simulated sequence of 2000 heartbeat periods in NSR, $RR_{NSR}$ (thin line), and during AF, $RR_{AF}$ (thick line). (b) PDFs of $RR_{NSR}$ (thin line) and $RR_{AF}$ (thick line). Reference thresholds are individuated by the $5th$ and $95th$ percentiles in NSR, while in AF they correspond to the $33rd$ and $72nd$ percentiles, respectively.

\noindent \textbf{Figure 3.} PV loops of the 2000 $RR$ beats: (a) NSR, (b): AF.

\noindent \textbf{Figure 4.} PDFs of systolic (thicker line) and diastolic (thinner line) pressures at the indicated sites in NSR (dotted line) and AF (continuous line). Representative pressure time-series are reported for NSR (dotted line) and AF (continuous line). $x$ [cm] is the distance from the heart ($x=0$ is the beginning of the ascending aorta).

\noindent \textbf{Figure 5.} (a) Mean and (b) $cv$ values for systolic (triangles), diastolic (bullets) and pulse (squares) pressures, along the arterial tree in NSR. $x$ [cm] is the distance from the heart.

\noindent \textbf{Figure 6.} (a)-(b) $cv$ ratios between AF and NSR along the arterial tree for systolic (triangles), diastolic (bullets), and pulse (squares) pressures. Horizontal lines indicate the $cv$ ratios at the ascending aorta. (c)-(d)-(e) Percentile variations in AF ('+': AF percentile corresponding to the $5th$ in NSR; '*': AF percentile corresponding to the $95th$ in NSR) along the arterial pathway, $x$, for (c) systolic, (d) diastolic, and (e) pulse pressures. Horizontal lines display the $5th$ and $95th$ percentiles in NSR.

\noindent \textbf{Figure 7.} (a) Mean and (b) $cv$ values for the phase velocity, $c$, along the arterial tree in NSR. (c) $cv$ ratio between AF and NSR along the arterial tree, $x$. The dotted line marks the $cv$ ratio for the ascending aorta. (d) Percentile variations of the phase velocity, $c$, in AF ('+': AF percentile corresponding to the $5th$ in NSR; '*': AF percentile corresponding to the $95th$ in NSR) along $x$. Horizontal lines show the $5th$ and $95th$ percentiles in NSR.

\noindent \textbf{Figure 8.} (a) $cv$ ratio between AF and NSR for $RM$ ($\rhd$) along the arterial tree, $x$. The dotted line marks the $cv$ ratio at the ascending aorta. (b) AF percentile variations for $RM$ ('+': AF percentile corresponding to the $5th$ in NSR; '*': AF percentile corresponding to the $95th$ in NSR). Horizontal lines report the $5th$ and $95th$ percentiles in NSR.

\section*{Table Legends}

\noindent \textbf{Table 1.} Statistics of the $RR$ sequence (2000 beats) in NSR ($RR_{NSR}$) and AF ($RR_{AF}$). $\mu$: mean value, $\sigma$: standard deviation, $cv$: coefficient of variation, $S$: skewness, $K$: kurtosis.

\noindent \textbf{Table 2.} Mean value, $\mu$, $\pm$ standard deviation, $\sigma$, of the indicated cardiac parameters and aortic pressures in both NSR (II Column) and AF (III Column). The IV Column gives the $cv$ ratios between AF and NSR, $cv_{AF}/cv_{NSR}$. Average variations during AF with respect to the NSR are reported in the V Column (together with the p-values of the t-test between NSR and AF), while the corresponding available literature variations are indicated in the VI Column. "+" increase during AF, "-" decrease during AF, "=" no significant variations during AF, "/" no data available. $EDV$: end-diastolic left-ventricular volume, $ESV$: end-systolic left-ventricular volume, $EDP$: end-diastolic left-ventricular pressure, $ESP$: end-systolic left-ventricular pressure, $SV$: stroke volume, $CO$: cardiac output, $EF$: ejection fraction, $SW$: stroke work, $P_{AA,sys}$, $P_{AA,dia}$ and $PP_{AA}$: systolic, diastolic and pulse pressures at the entrance of the aorta, respectively.

\noindent \textbf{Table 3.} Mean flow rates per beat, $\overline{Q}$ [ml/beat], at different abdominal regions along the arterial tree (numeration according to Fig. 1). $cv$ ratios between AF and NSR (II column), and AF percentile variations corresponding to the $5th$ and $95th$ percentiles in NSR (III and IV columns).

\clearpage

\clearpage

\begin{table}
\centering
\caption{}
\label{tab:1}
\begin{tabular}{cccccc}
\hline\noalign{\smallskip}
Case & $\mu$ [s] & $\sigma$ [s] & $cv$ & $S$ & $K$\\
\noalign{\smallskip}\hline\noalign{\smallskip}
\\
$RR_{NSR}$ & 0.80 & 0.06 & 0.07 & -0.08 & 3.06\\
$RR_{AF}$ & 0.80 & 0.19 & 0.24 & 0.75 & 4.73\\
\noalign{\smallskip}\hline
\end{tabular}
\end{table}

\begin{table}
\caption{}
\label{tab:2}
\begin{tabular}{cccclc}
\hline\noalign{\smallskip}

\multirow{2}{*}{Variable} &  \multicolumn{2}{c}{$\mu\pm\sigma$}  & \multirow{2}{*}{$\frac{cv_{AF}}{cv_{NSR}}$} & \multicolumn{2}{c}{Mean variations in AF with respect to NSR} \\
         &     NSR         &     AF          &                            &  Present results    &   Available literature    \\
\noalign{\smallskip}\hline\noalign{\smallskip}
\\
$EDV$ [ml] & 120.21 $\pm$ 0.04 & 120.09 $\pm$ 1.17 & 31.41 & - \,\, (p = 5 $\cdot 10^{-6}$) & = \cite{orlando}, + \cite{anter,therkelsen}\\
$ESV$ [ml] & 53.64 $\pm$ 1.23 & 54.18 $\pm$ 3.70 & 2.97 & + \, (p = 7 $\cdot 10^{-10}$) & = \cite{orlando}, + \cite{anter,therkelsen}\\
$EDP$ [mmHg] & 8.35 $\pm$ 0.01 & 8.34 $\pm$ 0.09 & 31.42 & - \,\, (p = 9 $\cdot 10^{-7}$) & - \cite{alboni}\\
$ESP$ [mmHg] & 93.27 $\pm$ 2.29 & 92.26 $\pm$ 7.59 & 3.35 & - \,\, (p = 1 $\cdot 10^{-8}$) & /\\
$SV$ [ml] & 66.57 $\pm$ 1.27 & 65.91 $\pm$ 4.19 & 3.34 & - \,\, (p = 2 $\cdot 10^{-11}$) & - \cite{alboni,corliss,giglioli,halmos,samet}, = \cite{killip}\\
$CO$ [l/min] & 4.99 $\pm$ 0.09 & 4.94 $\pm$ 0.31 & 3.34 & - \,\, (p = 6 $\cdot 10^{-12}$) & - \cite{clark,corliss,daoud,halmos,khaja,morris,orlando,samet,upshaw},
\\
 & & & & &  = \cite{resnekov,shapiro}\\
$EF$ [\%] & 55.38 $\pm$ 1.04 & 54.87 $\pm$ 3.30 & 3.20 & - \,\, (p = 5 $\cdot 10^{-11}$) & - \cite{anter,chirillo,gentlesk,therkelsen,wozakowska-kaplon}\\
$SW$ [J] & 0.92 $\pm$ 0.01 & 0.91 $\pm$ 0.04 & 3.10 & - \,\, (p = 0) & - \cite{khaja,orlando}\\
$P_{AA,sys}$ [mmHg] & 120.50 $\pm$ 3.17 & 120.73 $\pm$ 9.34 & 2.94 & + \, (p = 3 $\cdot 10^{-1}$) & + \cite{giglioli,kaliujnaya}, = \cite{alboni,clark}\\
$P_{AA,dia}$ [mmHg] & 71.00 $\pm$ 3.04 & 71.64 $\pm$ 9.69 & 3.16 & + \, (p = 5 $\cdot 10^{-3}$) & + \cite{alboni,giglioli,kaliujnaya}, = \cite{clark} \\
$PP_{AA}$ [mmHg] & 49.51 $\pm$ 0.31 & 49.10 $\pm$ 1.40 & 4.51 & - \,\, (p = 0) & - \cite{kaliujnaya}\\
\noalign{\smallskip}\hline
\end{tabular}
\end{table}

\begin{table}
\caption{}
\label{tab:3}
\begin{tabular}{cccc}
\hline\noalign{\smallskip}
\multirow{2}{*}{Vessel} & \multirow{2}{*}{$\frac{cv_{AF}}{cv_{NSR}}$} & AF percentile corresponding & AF percentile corresponding \\
         &             &     to the $5th$ percentile in NSR         &          to the $95th$ percentile in NSR                     \\
\noalign{\smallskip}\hline\noalign{\smallskip}
\\

Ascending aorta (1) & 3.70 & 30.1 & 65.5 \\
Intercostal (26) & 2.26 & 30.6 & 69.3 \\
Celiac (29) & 1.75 & 28.2 & 68.1 \\
Superior mesenteric (34) & 1.79 & 28.1 & 70.7 \\
Left Renal (36) & 1.65	& 29.6 & 67.7 \\
Right Renal (38) & 1.67 & 28.2 & 66.9 \\
Inferior mesenteric (40) & 1.95 & 28.8 & 63.9 \\
\noalign{\smallskip}\hline
\end{tabular}
\end{table}

\end{document}